\documentclass{ws-procs961x669}            % default, citations in superscript
\begin{document}
\title{Displacement memory and BMS symmetries}

\author{Shailesh Kumar$^*$}

\address{Indian Institute of Information Technology Allahabad,\\
Devghat, Jhalwa, Uttar Pradesh-211015, India.\\
$^*$E-mail: shaileshkumar.1770@gmail.com}

%\author{Arpan Bhattacharyya}

%\address{Indian Institute of Technology Gandhinagar,\\
% Gujarat-382355, India\\
%E-mail: bhattacharyya.arpan@yahoo.com}

\begin{abstract}
This article reviews one of the most intriguing properties of black hole spacetimes known in the literature- \textit{gravitational memory effect}, and its connection with asymptotic symmetries, also termed as Bondi-van der Burg-Metzner-Sachs (BMS) symmetries, emerging near the horizon of black holes. Gravitational memory is a non-oscillatory part of the gravitational wave amplitude which generates a permanent displacement for freely falling test particles or test detectors. We highlight a model scenario where asymptotic symmetries appear as a soldering freedom in the context of stitching of two black hole spacetimes, and examine the impact of the interaction between test detectors and horizon shells. Further, we provide a more realistic approach of computing displacement memory for near-horizon asymptotic symmetries which is analogous to the conventional memory originally obtained at asymptotic null infinity. %We also touch upon the possibility of investigating the signatures of asymptotic symmetries in other possible directions such as gravitational lensing and quasinormal modes which might play a significant role in understanding the BMS memory not only from theoretical perspectives but also from observational standpoints. 
\end{abstract}


\bodymatter

\section{Introduction}\label{aba:sec1}

The observational facets of gravitational waves (GWs) \citep{PhysRevLett.116.061102, PhysRevLett.125.101102} have opened a new window to look for various aspects of black hole spacetimes; \textit{gravitational memory}\citep{Zeldovich:1974gvh, Braginsky:1985vlg, Christodoulou:1991cr, PhysRevD.89.084039, Strominger:2014pwa, PhysRevD.101.124010} is one of such intriguing features that has not been detected yet. GW induces a permanent relative change in the position of test detectors by imparting a memory to the configuration. This permanent relative change is referred to as gravitational memory. The term \textit{memory} implies the information or properties of spacetimes from where it is being generated and carried by gravitational waves. The first practical computation of memory's evolution was done by Marc Favata using post-Newtonian formalism where he accounted for all stages of BBH coalescence\citep{Favata:2011qi, PhysRevD.80.024002, Favata_2009, Favata_2009n}. In this direction, recently, there have been several implications of detecting GW memory using advanced detectors\cite{Islo:2019qht, Lasky:2016knh, Boersma:2020gxx, Pollney_2011, Grant:2021hga, PhysRevD.101.023011, Islam:2021old, Hubner:2021amk}. 

On the other hand, it has been shown that the gravitational memory is closely related to the asymptotic symmetries of spacetimes originally discovered by Bondi-van der Burg-Metzner-Sachs (BMS) in the early sixties\cite{doi:10.1098/rspa.1962.0161}, and such symmetries can also be recovered near the horizon of black holes which motivated us to probe the near horizon properties of black holes. The recent findings in this direction have provided some strong grounds for the information loss puzzle. In the context of asymptotic symmetries, the existence of soft hair on black holes is necessary for charge conservation of supertranslation and superrotation \citep{strominger2018lectures, PhysRevLett.116.231301, Strominger:2014pwa, PhysRevD.96.064013}. As the conservation principles are derived from the long-distance behaviour of fields close to spatial infinity, the presence of black holes should have no effect on them. We know that the conserved charges can be expressed as bulk integrals over any Cauchy surface. A contribution from the future event horizon should be taken into account for conserved charges as future null infinity is no longer a Cauchy surface in the presence of a classical black hole. In this direction, Strominger and Hawking’s latest discovery uses the asymptotic symmetries of the BMS group to prove that information is not lost rather stored in something called as a soft particle. Therefore, soft hair or low-energy quantum excitations may be carried by a black hole and leak information when it evaporates. This brings a direct motivation for the emergence of asymptotic symmetries near the horizon of black holes from conservation perspectives. %Furthermore, Hawking, Perry and Strominger (HPS) have shown an explicit treatment of soft hair in terms of soft gravitons or photons and examined that at the future edge of horizon, the complete information about their quantum state is preserved on a holographic plate [101]. 

Let us understand how memory and BMS symmetries are inter-connected with each other. Classically, for a given spacetime geometry, BMS transformations produce an infinite class of spacetime metrics that are physically unique or distinct. Assume that BMS transformations act on a given metric $g_{\mu\nu} (x^{\mu})$ with $x^{\mu} = (x^{0},x^{i})$, i.e., one time and three spatial coordinates. Such an action on the metric generates a completely different metric.
\begin{align*}
g_{\mu\nu}(x^{\mu})  \xrightarrow{\text{BMS transformation}} \tilde{g}_{\mu\nu}(x^{\mu}) .
\end{align*}     
The metric $g_{\mu\nu}(x^{\mu})$ and $\tilde{g}_{\mu\nu}(x^{\mu})$ are distinct and this relative change implies the generation of \textit{GW memory}, and also motivates us to seek for a connection between memory and BMS symmetries.  This change can be understood in the following way- GWs generated from a black hole spacetime carrying information or properties in terms of BMS parameters would interact with the detector setup placed at the asymptotic null infinity, this would induce a permanent relative change in the initial configuration of the setup. A similar setup can also be considered at a place near the horizon of a black hole. A persistent effect similar to that of null-infinity may again be observed. It provides a physical meaning to the inter-connection between GW memory and asymptotic symmetries emerging near the horizon of black holes. Technically, $g_{\mu\nu}(x^{\mu})$ can be thought of as a metric of a given asymptotically flat spacetime and $\tilde{g}_{\mu\nu}(x^{\mu})$ is the resultant metric appears as a consequence of the interaction between GWs and detectors which implies a net relative change in the configuration and gives a definition to the memory. Briefly, if we have two nearby timelike geodesics or inertial detectors described by the tangent vector $T^{\mu}$ together with a deviation vector $s^{\mu}$, and let us position them at the future null infinity. The evolution of the deviation vector before and after the interaction with gravitational waves will be captured in the geodesic deviation equation (GDE), written as
\begin{align}
\frac{D^{2}s^{\mu}}{d\tau^{2}} = -R^{\mu}{}_{\delta\sigma\lambda}T^{\delta}T^{\lambda}s^{\sigma}. \label{gde}
\end{align}
The solution of the GDE will give us a permanent relative change in the displacement vector $s^{\mu}$ which can further be related to supertranslation and will implicate the achievement of \textit{BMS displacement memory effect}. Our study provides an analogous effect and its connection with asymptotic symmetries for near the horizon of black holes. 
%\begin{figure}[h!]\centering
%    \includegraphics[height=8.0cm, width=7.5cm]{Minkowski.pdf} 
%    \caption{An schematic diagram showing two detectors $d_{1}$ and $d_{2}$ stationed at future null infinity will capture the memory effect due to the interaction between detector setup and gravitational waves generated at late time}\label{solderingn}
%\end{figure}

%As mentioned earlier that the extensive analysis of recovering asymptotic symmetries at null infinity has got considerable attention for near the horizon of black holes due to its theoretical and observational importance in information loss paradox. The study of near-horizon BMS symmetries had started in the new millennium \cite{PhysRevD.64.124012, Hotta_2001}
There are two methods to recover asymptotic symmetries near the horizon of a black hole. As a recent progress, Donnay et al. showed the first way of obtaining such symmetries near the horizon of a stationary black hole \cite{PhysRevLett.116.091101} with asymptotic form of the Killing vectors preserving the boundary conditions. It turns out that the near-horizon region of a stationary black hole spacetime induces supertranslations including semi-direct sum with extended asymptotic symmetry superrotations which is being represented by Virasoro algebra. Hence, one can recover asymptotic symmetries that would mimic the ones originally obtained at asymptotic null infinity \cite{PhysRevLett.116.091101, Donnay2016} by preserving the near-horizon asymptotic structure of black holes. The second method for recovering asymptotic symmetries deals with the soldering of two spacetimes across a common null hypersurface \cite{Blau:2015nee, PhysRevD.98.104009}. It has been shown that we can stitch them in infinite ways by demanding that the induced metric remains invariant under the translations generated by the null generators of the shell \cite{Blau:2015nee, PhysRevD.98.104009}. The freedom for the choice of the intrinsic coordinates on null hypersurface in the null-direction is termed as \textit{soldering freedom}, and also known as \textit{BMS-like soldering freedom}. Since these appear as a metric preserving transformations, hence, known as BMS-like symmetries or BMS-like transformations. We shall discuss the related details in section (\ref{2}).

The article is organized as follows. In section (\ref{2}), we discuss the intrinsic formulation of null shells placed at the horizon and how near-horizon asymptotic symmetries are recovered in the context of stitching of two black hole spacetimes. Further, in section (\ref{horizon}), we show how horizon shells carrying memory affect the displacement between two nearby test detectors or test particles for Schwarzschild and Extreme Reissner Nordstr$\ddot{o}$m (ERN) black holes. We have also studied the impact of interaction between null geodesics and horizon shells; since we shall be completely focusing on timelike geodesics in this article, we do not include the discussion on null geodesics crossing the horizon shells. However, the study can be found in\cite{PhysRevD.100.084010, PhysRevD.102.044041}. We further consider a more realistic approach in section (\ref{3}) for determining the displacement memory effect and its connection with near-horizon asymptotic symmetries. In the end, we conclude our findings in section (\ref{discuss}) by providing some remarks on possible future outlooks to our studies which might be relevant from theoretical as well as observational perspectives.

\section{Horizon shell and asymptotic symmetries}\label{2}

In general relativity, a shell is a geometric configuration that can be used to investigate the propagation of thin distribution of null matter (e.g. neutrino) and impulsive gravitational waves (IGWs). The thin surface layer of null matter together with impulsive waves is precisely referred to as \textit{thin-shell} or \textit{thin null shell}\citep{PhysRevD.43.1129, nullsurface}. The generated impulsive signals are usually produced during violent astrophysical phenomena like supernova explosions or coalescence of black holes. If we stitch two black hole spacetimes along a common null hypersurface which also happens to be the horizon of black holes, and stitching is consistent with \textit{junction conditions}, we obtain \textit{horizon shell}. The soldering formalism shows that the stess-energy tensor of the stitched spacetime satisfying Einstein field equation carries a singular term proportional to the Dirac delta distribution function, given by 
\begin{align}\label{et}
        T_{\mu\nu} = T^{+}_{\mu\nu}\mathcal{H}(\Phi)+T^{-}_{\mu\nu}\mathcal{H}(-\Phi)+S_{\mu\nu}\delta(\Phi),
\end{align}
where $\mathcal{H}(\Phi)$ is a Heaviside step function for a given null surface $\Sigma\equiv \Phi =0$. The last term of Eq.(\ref{et}) corresponds to the stress-energy tensor of the null surface, and further investigation on the same shows the generation of \textit{impulsive gravitational wave} or thin surface layer of null matter or a mixture of both. 
\begin{figure}[h!]\centering
    \includegraphics[width=8.0cm, height=2.5cm]{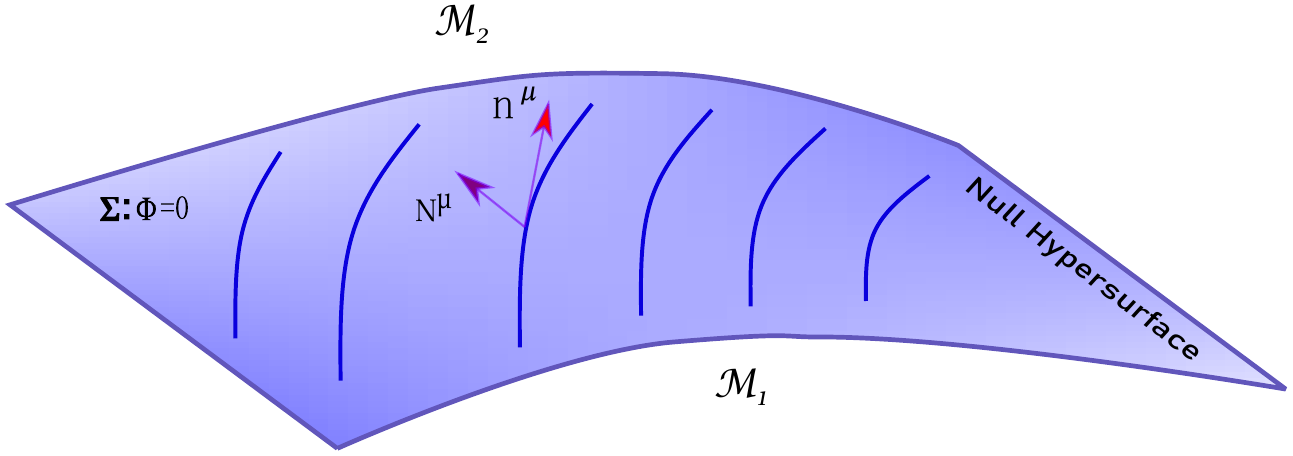} 
    \caption{Null hypersurface $\Sigma$ separating manifolds $\mathcal{M}_{1,-}$ and $\mathcal{M}_{2,+}$ each with a different metric.}\label{memff}
\end{figure}
The null hypersurface, representing the history of impulsive lightlike signals, separates spacetime manifold into two parts ($\mathcal{M}_{1}, \mathcal{M}_{2}$) or ($\mathcal{M}_{-}, \mathcal{M}_{+}$) as depicted in the Fig.(\ref{memff}) each with a different metric. $n^{\mu}$ is a null-normal or a generator of the null hypersurface, and $N^{\mu}$ is a transverse or auxiliary normal which is not tangent to the null surface, satisfying $n\cdot N=-1$. $N^{\mu}$ carries the transversal properties of $\Sigma$. One can study the intrinsic quantities of a horizon shell known as surface energy density ($\mu$), surface current ($J^{A}$) and surface pressure ($p$) in terms of \textit{transverse curvature} by analyzing the stress tensor on the null surface. The intrinsic quantities can be written in the following form
\begin{align}\label{intrinsic}
\mu = -\frac{1}{8\pi}\sigma^{AB}[\mathcal{K}_{AB}] \hspace{2mm} ; \hspace{2mm} j^{A} = \frac{1}{8\pi}\sigma^{AB}[\mathcal{K}_{V B}] \hspace{2mm} ; \hspace{2mm}
p = -\frac{1}{8\pi}[\mathcal{K}_{VV}],
\end{align}
where $`[\hspace{2mm}]'$ denotes the difference between a quantity computed on the null surface $\Sigma$ for both sides $\mathcal{M}_{1,2}$ separately. We notice that the null shell quantities depend on the induced metric $\sigma^{AB}$ and jump in the extrinsic curvature $\mathcal{K}_{AB}$\footnote{We have taken Kruskal-like coordinates. Also, Capital Latin letters denote spatial or spherical coordinates of the 2-sphere metric, and lower Latin letters denote hypersurface coordinates.}. The other related details can be found in \cite{nullsurface, Blau:2015nee, PhysRevD.100.084010, PhysRevD.102.044041}.

An induced tensor field $\gamma_{ab}$ on $\Sigma$ is related to the jump of the induced metric $g_{ab}$ and transverse normal $N^{\mu}$ which helps us to examine the intrinsic formulation of the horizon shell. The tensor field $\gamma_{ab}$ can be written in terms of the jump in the derivative of the induced metric along the auxiliary normal $N^{\mu}$ which can further be written in terms of transverse curvature, i.e., $\gamma_{ab}=N^{\mu}[\partial_{\mu}g_{ab}]=2[\mathcal{K}_{ab}]$. The analysis for intrinsic expression of stress tensor ensures that there is a part of $\gamma_{ab}$ which does not contribute to stress-energy tensor. We denote this non-contributing part as $\hat{\gamma}_{ab}$. In general, one can write down $\gamma_{ab}$ containing both null matter ($\bar{\gamma}_{ab}$) and GW degree of freedom ($\hat{\gamma}_{ab}$) in the following way
\begin{align}\label{q1}
\gamma_{ab} = \hat{\gamma}_{ab}+\bar{\gamma}_{ab},
\end{align} 
with
\begin{align}\label{q2}
\hat{\gamma}_{ab} =& \gamma_{ab}-\frac{1}{2}g_{*}^{cd}\gamma_{cd}g_{ab}-2n^{d}\gamma_{d(a}N_{b)}+\gamma^{\dagger}N_{a}N_{b} \\
\bar{\gamma}_{ab} =& 16\pi\Big(g_{ac}S^{cd}N_{d}N_{b}+g_{bc}S^{cd}N_{d}N_{a}-\frac{1}{2}g_{cd}S^{cd}N_{a}N_{b}-\frac{2}{2}g_{ab}S^{cd}N_{c}N_{d}\Big),
\end{align}
where $\gamma^{\dagger}=\gamma_{cd}n^{c}n^{d}$, and $g_{*}^{ab}$ is the pseudo-inverse of $g_{ab}$, i.e., $g_{*}^{ab}g_{bc}=\delta^{a}_{c}-n^{a}N_{c}, g_{*}^{cd}\gamma_{cd}=g^{AB}\gamma_{AB}$. The $\hat{\gamma}_{ab}$ carries the pure impulsive gravitational wave degree of freedom whereas $\bar{\gamma}_{ab}$ contains the null matter part of the horizon shell. In general, a null shell is being considered the combination of both IGWs and null matter. Further, using the expressions of intrinsic quantities of the null surface together with $\hat{\gamma}_{ab}$ and $\bar{\gamma}_{ab}$, in Kruskal coordinates, one obtains,
\begin{align}\label{q3}
\bar{\gamma}_{VB}=16\pi g_{BC}S^{VC} \hspace{2mm} ; \hspace{2mm} \bar{\gamma}_{AB} = -8\pi S^{VV}g_{AB}.
\end{align}
Now, we shall examine the interaction of such impulsive lightlike signals carrying BMS parameters on timelike geodesics. Let us first investigate the appearance of BMS symmetries in gluing formalism. 

\subsection{Emergence of asymptotic symmetries}

Here, we investigate the emergence of near-horizon asymptotic symmetries in the context of soldering of two black hole spacetimes. This, in gluing formalism, can be achieved via obtaining the freedom in the choice of intrinsic coordinates along the null direction.  The soldering freedom of stitching the two spacetimes along a common null surface provides BMS-like transformation on the horizon shell. It emerges as a coordinate transformation which preserves the induced metric on the null hypersurface $\Sigma$. This implies us to figuring out the Killing vectors of the hypersurface metric in a suitable coordinate system\citep{Blau:2015nee,PhysRevD.98.104009}. Therefore, the analysis is based on the Lie derivative of the induced metric along the Killing direction (say $Z^{a}\partial_{a}$ with components $Z^{a}$). We consider Kruskal coordinates $(U, V, x^{A})$ with coordinates $(V, x^{A})$ on $\Sigma$. Further, we also consider the metric with $g_{aV}=0$. Therefore, the Killing equation for spatial metric $g_{AB}$ is
\begin{align}\label{kilspa}
\mathcal{L}_{Z}g_{AB} = 0 \hspace{0.1cm} \Longrightarrow \hspace{0.1cm} Z^{V}\partial_{V}g_{AB}+Z^{C}\partial_{C}g_{AB}+(\partial_{A}Z^{C})g_{CB}+(\partial_{B}Z^{C})g_{AC} = 0.
\end{align}
Now, we may separately examine the emergence of near-horizon BMS symmetries- supertranslation and superrotation.

\subsubsection{Supertranslation}

The first special case is when metric does not depend on $V$ parameter, i.e., $\partial_{V}g_{AB}=0$. This induces a new type of translation which has angle dependent notion, termed as \textit{Supertranslation}. It is similar to the one obtained at asymptotic null infinity for asymptotically flat spacetimes. Keeping in mind the impact of the $Z$-generated transformations on the null normal $n^{a}$ of $\Sigma$, as a result, Eq.(\ref{kilspa}) gives
\begin{align}\label{supt}
\partial_{V}Z^{V} = 0 \hspace{0.3cm} \Longrightarrow \hspace{0.3cm} V \hspace{0.2cm} \longrightarrow \hspace{0.2cm} V+T(x^{A}) \hspace{0.2cm} : \hspace{0.2cm} BMS \hspace{0.15cm} Supertranslation
\end{align} 
This is instantly indentified as a supertranslation in the literature with $T(x^{A})$ being a supertranslation parameter, where $x^{A}=(\theta, \phi)$. It is interpreted as an angle dependent translation, hence named \textit{supertranslation}. The soldering group that keeps this structure preserved is still infinite dimensional. Let us now turn our discussion to investigate the extended form of asymptotic symmetry.

\subsubsection{Superrotation}\label{suerr}

A new type of symmetry labeled as \textit{superrotation} has just been discovered in an extended form of BMS symmetries near the horizon of black holes which mimics the one obtained at asymptotic null infinity \cite{PhysRevLett.105.111103, barnich2012supertranslations, PhysRevLett.116.091101, Blau:2015nee, PhysRevD.98.104009}. It is a local conformal transformation of the spatial slice of the metric, or local conformal transformation of celestial sphere at null infinity \cite{strominger2018lectures, PhysRevLett.105.111103}. Let us determine the extended BMS symmetry by considering the case when the spatial slice of the metric depends on the $V$ parameter, i.e., $\partial_{V}g_{AB}\neq 0$. The analysis begins with the Eq.(\ref{kilspa}) in search of possible non-trivial soldering freedoms. If one performs the conformal transformation in spatial coordinates represented in complex coordinates via $z \longrightarrow f(z)$ and $\bar{z} \longrightarrow \bar{f}(\bar{z})$ such that the Eq.(\ref{kilspa}) can be written in the following way
\begin{align}\label{suprreq}
Z^{V}\partial_{V}g_{AB} + \Omega(x^{A})g_{AB} = 0,
\end{align}
where $x^{A}=(z, \bar{z})$ and $\Omega(x^{A})$ denotes the conformal factor. We have the equation whose feasible solution can be written as, $g_{AB} = r^{2}(U,V)\tilde{g}_{AB}(x^{A})$. It gives a suitable choice along $V$ direction, i.e., $Z^{V}(=-\frac{V\Omega(x^{A})}{2})$  which compensates the conformal transformation. This ensures that the metric remains preserved under such transformations. Thus, the analysis gives rise metric preserving extended BMS transformations known as \textit{superrotation}-like symmetries.

\section{Displacement memory and asymptotic symmetries}

As a first approach, we wish to examine the relative change in the displacement vector between two nearby timelike geodesics which arises due to interaction with impulsive lightlike signals. We also discuss the appearance of near-horizon asymptotic symmetries in the context of soldering of two ERN and Schwarzschild spacetimes. Second, we would be considering a more realistic approach of computing displacement memory which is analogous to the one obtained at asymptotic null infinity. Let us first start with the case where horizon shell interacts with test detectors.

\subsection{Memory \& BMS symmetries due to impulsive lightlike signals}\label{horizon}

Cataclysmic processes such as black hole mergers and supernovae explosions produce shockwave type of gravitational radiations. We wish to estimate the finite difference in the displacement vector between two nearby timelike test particles or geodesics upon crossing the horizon shell\footnote{The study ($B$-memory) of interaction between null congruence and horizon shell can be found in Bhattacharjee et al\cite{PhysRevD.100.084010, PhysRevD.102.044041}.}. It turns out that the asymptotic symmetries, associated with impulsive lightlike signals, leave footprints on test particles upon passing through them. As a result, we studied the effects of horizon shell for Schwarzschild and extreme RN spacetimes on the separation vector of two nearby timelike geodesics. The displacement between the geodesics is identified by supertranslation parameter which gives us \textit{Supertranslation memory effect}. Let us consider a congruence having $T^{\mu}$ to be a tangent vector with $T\cdot T=-1$. A displacement or separation vector between two test particles is $X^{\mu}$ satisfying $T\cdot X=0$. Thus one can compute the relative change in the separation vector before and the passage of impulsive lightlike signals by analyzing the GDE Eq.(\ref{gde}). %, i.e.,
%\begin{align}
%\ddot{X}^{\mu}=-R^{\mu}{}_{\alpha\beta\gamma}T^{\alpha}X^{\beta}T^{\gamma},
%\end{align}  
Riemann tensor is the memory generating factor for the given configuration. The solution of the GDE will generate a non-vanishing finite change in the deviation vector upon interacting with IGWs. Following the basic framework of the analyses from \cite{PhysRevD.43.1129, PhysRevD.100.084010, PhysRevD.102.044041}, we use the expressions written in section(\ref{2}) and $X^{a}=\tilde{g}^{ab}X_{b}$, we obtain components of deviation vectors as
\begin{align}
X_{V} =& 8\pi Ug_{BC}S^{VC}X^{B}_{(0)} \\
X_{A} =& X_{A(0)}+\frac{U}{2}\gamma_{AB}X_{(0)}^{B}+UV^{-}_{(0)A},
\end{align} 
where $V^{-}_{(0)a}=\frac{dX^{-}_{a}}{dU}\Big\vert_{U=0}$, and $\tilde{g}_{ab}=g_{ab}+(T_{(0)\mu}e^{\mu}_{a})(T_{(0)\nu}e^{\nu}_{b})$ with $e_{a}$ defined as a triad on the null surface. The $X_{(0)}^{B}$ is some function evaluated on the null surface, and it is denoted by subscript $(0)$. It is to note that when $S^{VC}$ is nonzero, then we have $X_{V}\neq 0$. This implies that the particle will be displaced off from the initial two dimensional surface. On the other hand, if $S^{VC}=0$, the component $X_{V}$ vanishes which means that the particle will reside on the initial two dimensional surface but with a relative displacement. In this particular consideration, the nonzero displacement vector $X_{A}$ is written as
\begin{align}
X_{A} =& (1-4\pi US^{VV})\Big(g_{AB}+\frac{U}{2}\hat{\gamma}_{AB}\Big)X^{B}_{(0)}.
\end{align}
The factor $\gamma_{AB}$ is the one which carries the BMS memory part of the wave, and generates the distortion effect on the test particles. This sets our first goal to investigate the memory signal arises in the context of soldering of two black hole spacetimes. Further, we show our studies for extreme RN and Schwarzschild black holes. Let us understand these two cases separately.

\subsubsection{Extreme RN case (ERN)} 
As we know that $70\%$ astrophysical black holes are near extremal and many super-massive black holes are also near extremal \cite{Volonteri_2005,Gou_2014,McClintock_2006,Brenneman_2006}. Further, Strominger and Vafa \cite{Strominger1996} determined the Bekenstein-Hawking area-entropy relation for extreme black hole. So extreme black holes are important from experimental as well as theoretical perspectives. Here, we investigate the asymptotic symmetries together with intrinsic properties of the shell and its interaction with test detectors in terms of BMS parameters. It is known that Carter investigated the maximal analytic extension of RN black hole for $e^{2} = M^{2}$ \cite{CARTER1966423}. As the Carter's metric is not $C^{1}$ i.e. the first derivative of the metric component is discontinuous. Secondly, he certainly did a conceptual analysis without providing the exact Kruskal analogue for the extreme case. For our purpose, it is important to have an exact form of the Kruskal metric which enables us to write $U = 0$ on the horizon. This helps us to perform off-shell extension of the soldering transformation without any obscurities. Therefore, We adopt a Kruskal extension that unambiguously places the shell at $U = 0$, and also better suited for memory effect. The ERN metric in Kruskal coordinates can be written as \citep{PhysRevD.62.024005}
\begin{align}\label{rnm}
ds^{2} = -\frac{2M}{r^{2}}\psi(V)'dU dV+r^{2}(U)(d\theta^{2}+\sin^{2}\theta d\phi^{2}),
\end{align}
where, $\psi(V)'$ is a regular, defined as, $\psi(V)=4M\Big(lnV-\frac{M}{2V}\Big)$. Also $U=-(r-M)$ where $r=M$ is the horizon.

Immediately, by looking at the spherical part of the metric, we observe that supertranslation-like symmetries can be recovered, written as: $V\longrightarrow V+T(\theta,\phi)$. We also find that coordinate $r$ is independent of $V$, thus interestingly, ERN consideration does not induce the superrotation-like symmetries whereas it is not the case with Schwarzschild discussed in section(\ref{sch}). For explicit details, we refer to Bhattacharjee et al\citep{PhysRevD.102.044041}.

\begin{figure}[h]
 \centering
 \includegraphics[scale=0.6]{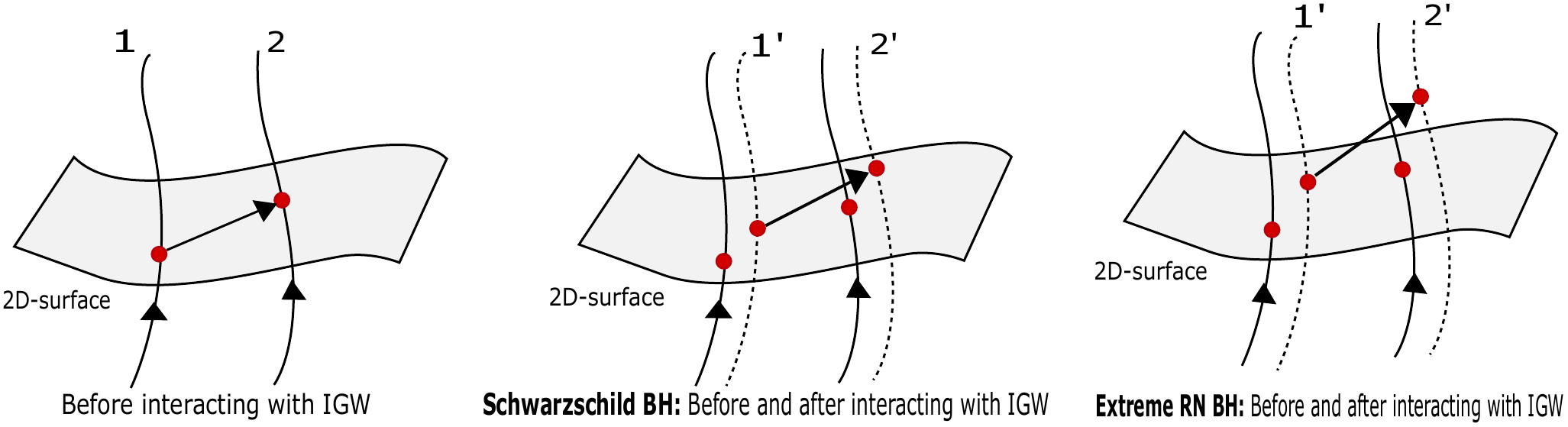}
 \caption{Timelike geodesics 1 \& 2 get displaced upon interacting with IGW, depicted as 1' \& 2' with a new relative displacement vector.}\label{displacmenttime}
 \end{figure}

Further, we examine some measurable effects on timelike test particles due to interaction with horizon shell of ERN spacetime. In this process, we first extend the soldering transformation off the horizon shell to the linear order in $U$. The transformations are give by\citep{PhysRevD.100.084010}
\begin{align}\label{trns}
U_{+} = UC(V,x^{A}) \hspace{2mm} ; \hspace{2mm} V_{+}=F(V,x^{A})+UA(V,x^{A}) \hspace{2mm} ; \hspace{2mm} x_{+}^{A}=x^{A}+UB^{A}(V,x^{A}),
\end{align}
where $x^{A}\equiv (\theta, \phi)$, with
\begin{align}
C = \frac{\partial_{V}\psi(V)}{\partial_{V}\psi(F)} \hspace{2mm} ; \hspace{2mm} A = \frac{M^{2}}{2}\frac{F_{V}}{\partial_{V}\psi(V)}\sigma_{AB}B^{A}B^{B} \hspace{2mm} ; \hspace{2mm} B^{A} = \partial_{V}\psi(V)\frac{1}{M^{2}F_{V}}\sigma^{AB}F_{B} .
\end{align}
Here, $\sigma_{AB}$ denotes the unit 2-sphere metric. One side of the spacetime $\mathcal{M}_{-}$ is completely ERN and the another side of the spacetime $\mathcal{M}_{+}$ is off-shell extended with the transformations (\ref{trns}). The process of obtaining the intrinsic quantities is known as \textit{off-shell extension of the soldering transformations}. It is to note here that one can also obtain the intrinsic properties of the horizon shell using \textit{extrinsic curvature algorithm}. Both the results would exactly match. The benefit of the later approach is that it makes the computational algebra significantly simplified. As a result, we find that the surface current ($j^{A}$) together with surface energy density ($\mu$) and pressure ($p$) is nonvanishing, and can be expressed in terms of supertranslation parameter $T(\theta,\phi)$. For example, the surface current is given by
\begin{align}
j^{A} = \frac{1}{8M^{2}\pi} \sigma^{AB}\Big(T_{B}\frac{\psi(T)''}{\psi(T)'}\Big).
\end{align}
The expressions for $\mu$ and $p$ can also be found in \cite{PhysRevD.102.044041}. The presence of nonzero surface current induces a finite change in the $X_{V}$ component of the deviation vector, i.e., $X_{V}\neq 0$ ; therefore, test particles get displaced off the initial 2-dimensional surface with a relative change in the displacement vector. The $X_{\theta}$ component is given by%of the deviation vector can be written in the following way
\begin{align}
X_{\theta} = X_{\theta (0)}+\frac{U}{2}\Big(\gamma_{\theta\theta}X^{\theta}_{(0)}+\gamma_{\theta\phi}X^{\phi}_{(0)}\Big)+UV^{-}_{(0)\theta},
\end{align}
where,
\begin{align}
\gamma_{\theta\theta} =& 2\psi(V)'\Big(T_{\theta\theta}+\frac{T_{\theta}^{2}\psi(T)''}{\psi(T)'}-\frac{M}{\psi(T)'}+\frac{M}{\psi(V)'}\Big)\\
\gamma_{\theta\phi} =& \gamma_{\phi\theta} = 2\psi(V)'\Big(\frac{T_{\theta}T_{\phi}}{\psi(T)'}\psi(T)''+T_{\theta\phi}-T_{\phi}\cot\theta\Big) .
\end{align}
This helps in determining GW degree of freedom $\gamma_{ab}$. One can determine $X_{\phi}$ component in a similar way. We notice that the deviation is written in terms of supertranslation parameter $T(\theta,\phi)$. The integration with respect to the geodesic parameters would give rise the displacement memory which would mimic the one obtained at null infinity. The Fig.(\ref{displacmenttime}) depicts the ultimate result of the test particles getting displaced off from the initial spatial slice with a comparison on Schwarzschild discussed in the section below. This completes our analyses of examining the role of near-horizon asymptotic symmetries on test particles upon interacting with impulsive lightlike signals together with the intrinsic properties of ERN horizon shell. 
%This completes our study on intrinsic formulation of horizon shell and its interaction with timelike test particles togther with the emergence of asymptotic symmetries in the context of soldering of two black hole spacetimes.

\subsubsection{Schwarzschild case} \label{sch}
Now, we start with the Schwarzschild spacetime in order to examine the intrinsic formulation of the horizon shell and study the interaction with test particles\cite{PhysRevD.100.084010}. Let us write down the metric in Kruskal coordinates,
\begin{align}
ds^{2} = -2G(r)dU dV+r^{2}(U,V)(d\theta^{2}+\sin^{2}\theta d\phi^{2}),
\end{align}
where, $G(r) = \frac{16M^{3}}{r}$ and $UV-\Big(\frac{r}{2M}-1\Big)e^{r/2M}$.
The horizon is defined as $U=0$. One can clearly see the supertranslation-like transformation, written as $V\longrightarrow V+T(\theta,\phi)$. Further, we notice that spherical part of the metric can be written as $\gamma_{\zeta\bar{\zeta}}=r^{2}\frac{d\zeta d\bar{\zeta}}{(1+\zeta\bar{\zeta})^{2}}$, and given the conformal transformations as discussed in (\ref{suerr}), in contrast to ERN spacetime, we obtain superrotations for the null shell placed just outside the horizon $U=\epsilon $ with a compensation along null direction $V$ which makes sure that the soldering transformations also preserve the form of the metric. These transformations mimic the ones originally obtained at asymptotic null infinity. We also compute the intrinsic quantities of the horizon shell which contain supertranslation parameter $T(\theta,\phi)$, can be seen in Bhattacharjee et al\cite{PhysRevD.100.084010}. 

%\begin{align}
%\mu = -\frac{1}{32m^{2}\pi}(\Delta^{(2)}T(\theta,\phi)-T(\theta,\phi)) \hspace{2mm} ; \hspace{2mm} j^{A} = 0 \hspace{2mm} ; \hspace{2mm} p = 0 .
%\end{align}
%This makes sure that the energy density $\mu$ is conserved, i.e., $\partial_{V}\mu=0$.
Next, we show the non-vanishing finite change in the components of the deviation or displacement vector between two nearby timelike geodesics upon passing through the horizon shell. We follow Blau et al. \cite{Blau:2015nee} to extend the soldering transformations off the horizon shell in order to compute the induced tensor field $\gamma_{ab}$. The computation of the deviation vectors further require the GW degree of freedom which can be expressed as
\begin{align}
\hat{\gamma}_{\theta\phi} = 2\nabla_{\theta}^{(2)}\partial_{\phi}T(\theta,\phi) \hspace{2mm}; \hspace{2mm}
\hat{\gamma}_{\theta\theta} = 2\Big(\nabla^{(2)}_{\theta}\partial_{\theta}T(\theta,\phi)-\frac{1}{\sin^{2}\theta}\nabla_{\phi}^{(2)}\partial_{\phi}T(\theta,\phi)\Big),
\end{align}
where $T(\theta,\phi)$ is a supertranslation parameter. Thus the framework of section(\ref{2}) and section(\ref{horizon}) generates the $\theta$-component of the deviation vector,
\begin{align}\label{xt1}
X_{\theta} =& \Big(1+\frac{U}{8m^{2}}(\nabla^{2}T(\theta,\phi)-T(\theta,\phi)\Big)\Big(\Big(4m^{2}+U(\nabla^{(2)}\partial_{\theta}T(\theta,\phi) \nonumber\\ 
& -\frac{1}{\sin^{2}\theta}\nabla_{\phi}^{(2)}\partial_{\phi}T(\theta,\phi))\Big)X^{\theta}_{(0)}+U\nabla_{\theta}^{(2)}\partial_{\phi}T(\theta,\phi)X^{\phi}_{(0)}\Big).
\end{align}
The $X_{\phi}$ component can also be computed in the similar way which again carries supertranslation parameter. It turns out that the surface current vanishes, i.e., $J^{A}=0$, hence the test particles will remain on the spatial slice of the metric since $X_{V}=0$, but with a relative change in the displacement as it can clearly be seen in Fig.(\ref{displacmenttime}), and opposes the result of ERN spacetime. Therefore, the nonvanishing displacement vector $X_{A}$ depicts the BMS displacement memory in the context of soldering of two Schwarzschild spacetime geometries. One can further integrate Eq.(\ref{xt1}) with respect to the parameter of the geodesics in order to have the explicit form of the displacement memory. Next, we shall discuss a more realistic approach of estimating the displacement memory which is analogous to the far region analysis. 

%We have also considered the study on null congruence crossing a null surface orthogonally which gives rise the changes in optical parameters like shear, expansion, etc. We call it as $B$-memory as it is related to $B$-tensor where geodesics fails to remain parallel. Since in this article we highlight on the displacement memory test detectors or timelike geodesics, we avoid the discussion of $B$-memory here. The related details can be found in Bhattacharjee et al\cite{PhysRevD.100.084010}. %We estimate these changes for a congruence defined by the tangent vector $\mathcal{N}$ as it crosses the horizon shell.

\subsection{Memory and BMS symmetries: Analogous to far region}\label{3}

In this section, we adopt a more realistic approach of computing displacement memory near the horizon of black holes and its possible connection with asymptotic symmetries. The analysis of this section is independent of section(\ref{2}), i.e., it is not based on soldering of black hole spacetimes. We study displacement memory for non-extremal (fixed temperature) and extremal (zero temperature) black holes\cite{Bhattacharjee2021}. This shows an analogous effect of conventional GW-memory which was originally established at asymptotic null infinity ($I^{+}$). The emergence of asymptotic symmetries near the horizon of black holes (not in the context of null shell formalism) has been established by \cite{barnich2012supertranslations, PhysRevLett.116.091101}. In this respect, we are interested in measuring the permanent relative change in the deviation or displacement of the test detectors induced due to the interaction with GW, and its connection with near-horizon BMS symmetries. %This establishes actual shift or displacement between the two test masses or detectors near the horizon of a black hole spacetime before and after the passage of gravitational waves. 
In this realistic approach, as it can be seen in the schematic diagram below, the detectors are being placed near the horizon of a black hole ($\mathcal{H}^{+}$), and we estimate a relative change in the deviation vector of configuration before and after the passage of GWs. The displacement vector $S^{\mu}$ between the detector setup or geodesics evolves according to GDE Eq.(\ref{gde}).  
\begin{figure}[h]
 \centering
 \includegraphics[height=4.7cm, width=4.5cm]{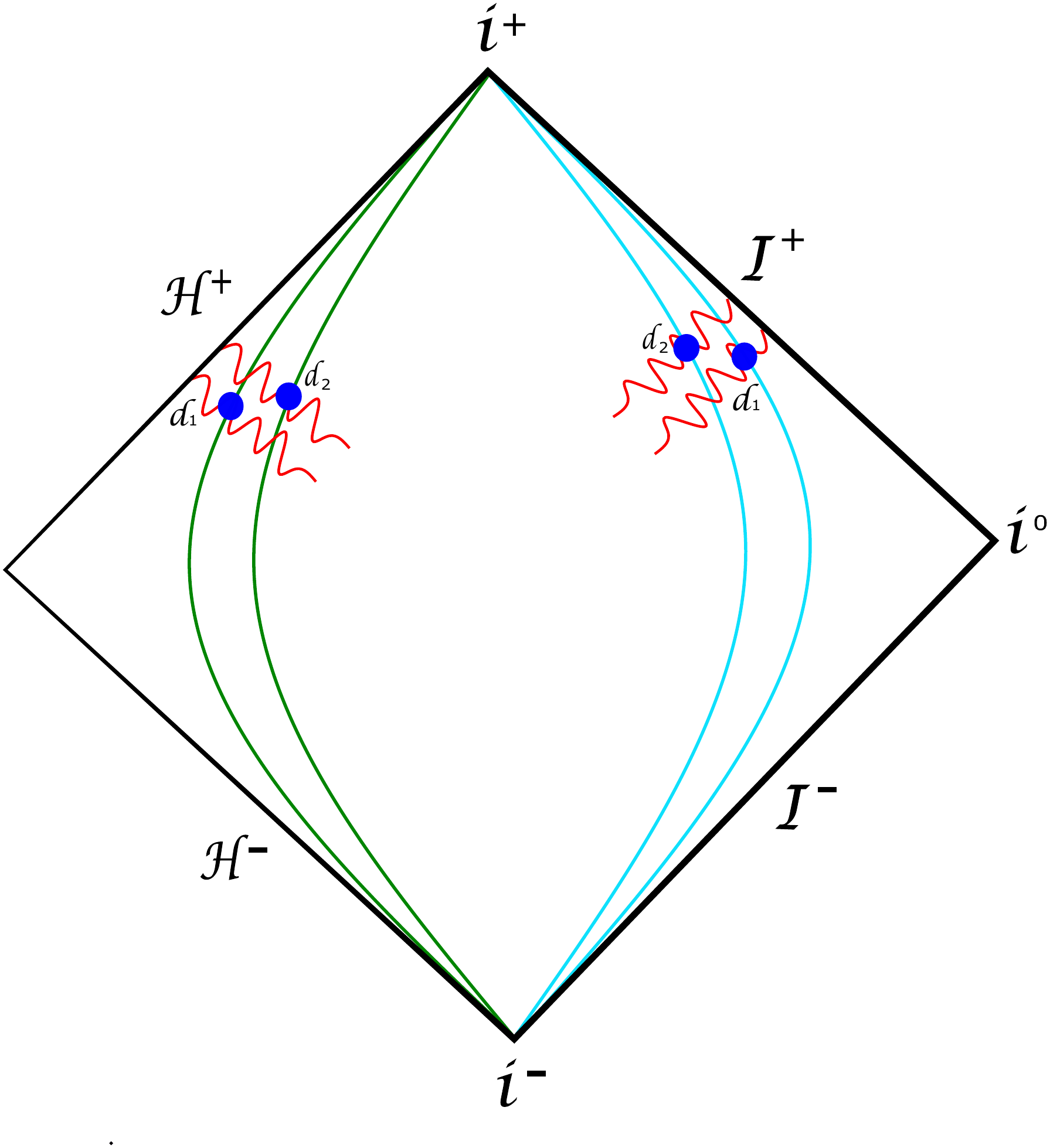} \qquad\qquad
 \includegraphics[height=4cm, width=5.5cm]{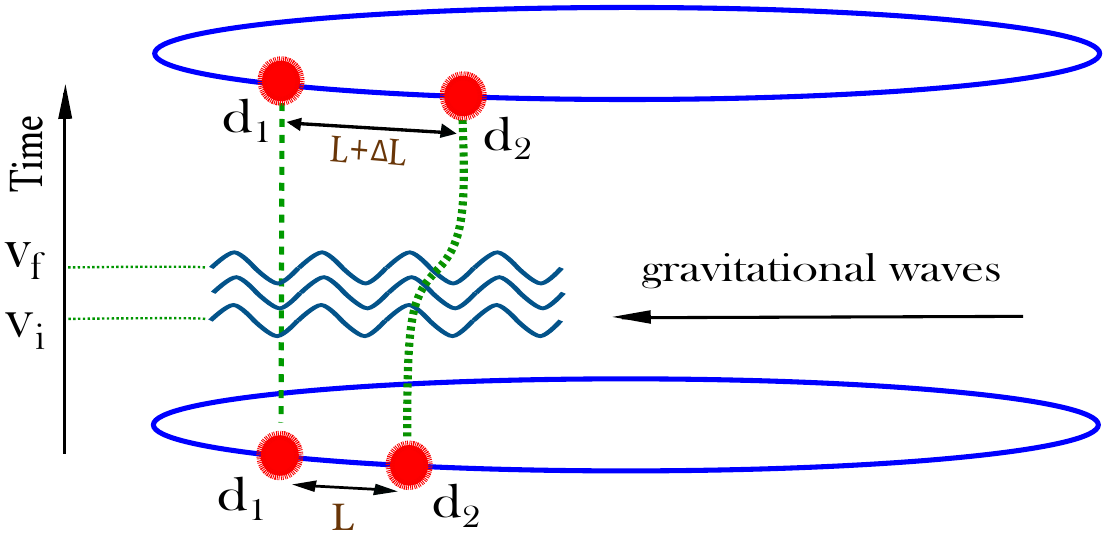}
  \caption{\footnotesize Schematic diagrams depicting displacement memory effect for the detectors $d_{1}$ and $d_{2}$. Separation $L$ gets modified permanently before and after the passage of GWs depicted as $L+\Delta L$.}
 \end{figure}
Let us consider the  general form of the 4-dimensional near-horizon metric \cite{Donnay2016,PhysRevD.98.124016}
\begin{align}
ds^{2} =& g_{vv} dv^{2}+2\kappa dv d\rho +2g_{vA} dv dx^{A}+g_{AB}dx^{A}dx^{B}, \label{m}
\end{align}
with following fall-off conditions for the horizon $\rho = 0 $:
\begin{align}
    g_{vv} =& -2\kappa\rho + \mathcal{O}(\rho^{2}) \hspace{3mm} ; \hspace{5mm} k = 1+\mathcal{O}(\rho^{2}) \nonumber\\
    g_{vA} =& \rho \theta_{A}+\mathcal{O}(\rho^{2}) \hspace{8.3mm} ; \hspace{5mm} g_{AB} = \Omega \gamma_{AB}+\rho \lambda_{AB}+\mathcal{O}(\rho^{2}) \nonumber
\end{align}
where $\theta_{A}$, $\Omega$ and $\lambda^{AB}$ are functions of ($v,x^{A}$). For computational purpose, we consider $\Omega$ to be unity. $\gamma_{AB}$ represents the 2-sphere metric. In stereographic coordinates, $x^{A}=(\zeta, \bar{\zeta})$, the 2-sphere metric is $\gamma_{AB} dx^{A} dx^{B} = \frac{4}{(1+\zeta\bar{\zeta})^{2}} d\zeta d\bar{\zeta}$. The asymptotic Killing vectors preserving fall-off boundary conditions together with the charges can be found in Donnay et al\cite{Donnay2016}. It turns out that the variation of $\kappa$ along Killing direction, when fixed temperature configuration considered, generates a copy of supertranslation together with a superrotation\cite{PhysRevLett.116.091101,Donnay2016}. We use this fact in section(\ref{bms}). Let us consider the fixed temperature configuration first in order to compute displacement memory.

\subsubsection{Memory: fixed temperature configuration}
%\textbf{\large Non-Extremal Case:}\\

The fixed temperature configuration enables us to take $\kappa$ non-zero but constant. With this consideation, the solution of the GDE for the given metric is\cite{Bhattacharjee2021}
\begin{align}
    \Delta S^{\bar{\zeta}} = \frac{\rho (1+\zeta\bar{\zeta})^{2}}{4}\Big((\Delta\lambda_{\zeta\bar{\zeta}}S^{\bar{\zeta}}+\Delta\lambda_{\zeta\zeta}S^{\zeta})-\kappa (\Delta v)^{2}(\mathcal{H}S^{\bar{\zeta}}+\mathcal{G}S^{\zeta})\Big)+\mathcal{O}(\rho^{2}), \label{ngd3n}
\end{align}
where $S^{\bar{\zeta}}$ and $S^{\zeta}$ denote the $\bar{\zeta}$ and $\zeta$ components of the deviation vector. Here, we have also used the $vA$-component of the Einstein field equation, the $\mathcal{O}(\rho^{0})$ term gives $\partial_{v}\theta_{A} = 0 \hspace{1mm} \Rightarrow \hspace{1mm} \theta_{A} = \mathcal{C}(x^{A})$. Using $vv$-component of the Einstein field equation, we replace changes $\Delta\lambda_{AB}$ in Eq.(\ref{ngd3n}) to obtain an explicit form of the memory which also ensures that $\lambda_{AB}$ can be written in terms of $\theta_{A}$. The resultant displacement memory is 
\begin{align}
    \Delta S^{\bar{\zeta}} = \frac{\rho (1+\zeta\bar{\zeta})^{2}}{4}\Big(\Big((\kappa\tilde{G}\Delta v-\frac{2aQ}{(1+\zeta\bar{\zeta})^{2}}) S^{\bar{\zeta}}+\kappa \tilde{B}S^{\zeta}\Big)\Delta v-\kappa (\Delta v)^{2}(\mathcal{H}S^{\bar{\zeta}}+\mathcal{G}S^{\zeta})\Big) \nonumber \\ +\mathcal{O}(\rho^{2}), \label{ngd3nn}
\end{align}
where $\tilde{G}, Q, \tilde{B}, \mathcal{H}$ and $\mathcal{G}$ are functions of ($\zeta,\bar{\zeta}$), also $\mathcal{H}$ is written in terms of metric parameter $\theta_{A}$. Thus the analyses suggest that the displacement memory is restored in terms of metric parameters. This completes our study of achieving the displacement memory near the horizon of non-extremal black holes. Further, we relate it with the asymptotic symmetries.

\paragraph{Relation with BMS symmetry:}\label{bms}
%\textbf{\large Relation with BMS Symmetry:}\\

We show the explicit relation between BMS symmetries and displacement memory. One can obtain the variation of the metric parameters $\lambda_{AB}$, $\theta_{A}$ and $\kappa $ along the Killing direction. Since the the memory (\ref{ngd3n}) or (\ref{ngd3nn}) is independent of $v$ coordinate, we only mention the relevant expression of $\lambda_{AB}$-variation\cite{Donnay2016}, given by  
\begin{align}
{\cal L}_{\chi}\lambda_{AB} = f \partial_{v}\lambda_{AB}-\lambda_{AB}\partial_{v}f+\mathcal{L}_{Y}\lambda_{AB}+\theta_{A}\partial_{B}f+\theta_{B}\partial_{A}f-2\nabla_{A}\nabla_{B}f . \label{chi2}
\end{align}
On the other hand, the $v$ component of the Killing vector for the fixed temperature configuration generates two sets of supertranslations $T(\phi)$ and $X(\phi)$, i.e.,
\begin{align}
    f(v,x^{A}) = T(x^{A})+e^{-\kappa v}X(x^{A}). \label{k4d}
\end{align}
Also, the Lie derivative of $g_{vA}$ along Killing direction yields superrotation $Y^{A}(x^{A})$\cite{Donnay2016}. 

Now, in order to make the variation of $\lambda_{AB}$ along the killing direction independent of $v$, we set $v$ coefficients to be zero. Using the general solutions of $\lambda_{AB}$ from relevant component of Einstein field equations, and for computational simplification switching off the supertranslation parameter $T$, we obtain $Y^{\zeta}(\zeta) = \tilde{a}e^{-\int\frac{\tilde{w}}{\tilde{p}}d\zeta}$ as a solution. where, $\tilde{w}$ and $\tilde{p}$ are functions of ($\zeta,\bar{\zeta}$) and $\tilde{a}$ appears as an integration constant; it would be a function of $\bar{\zeta}$ with respect to $\zeta$ differential equation. Similarly, one can also find the solution for $Y^{\bar{\zeta}}(\bar{\zeta})$. Therefore, we can find a solution for $Y^{A}$ that will induce the desired shift in the displacement vector, and serves our purpose of establishing relation between displacement memory and asymptotic symmetries near the horizon of black holes. 

%Whereas the general solutions for $\lambda_{AB}$ can be written as
%\begin{align}
%    \lambda_{\zeta\bar{\zeta}} = A(\zeta,\bar{\zeta})+B(\zeta,\bar{\zeta})v+\kappa\tilde{\mathcal{K}}(\zeta,\bar{\zeta})v\Delta v \hspace{1mm} ; \hspace{1mm} \lambda_{\zeta\zeta} = \tilde{A}(\zeta,\bar{\zeta})+v\tilde{B}(\zeta,\bar{\zeta}) \kappa \hspace{1mm};\hspace{1mm} \lambda_{\bar{\zeta}\bar{\zeta}}=\tilde{W}(\zeta,\bar{\zeta})+v\kappa\tilde{Z}(\zeta,\bar{\zeta}) \label{nd4}
%\end{align}

\subsubsection{Memory: Zero temperature configuration \& BMS symmetries}
%\textbf{\large Extremal Case:}\\

We have also provided an explicit approach for zero-temperature configuration (extremal consideration). For this, the metric coefficients remain same as appear in the non-extreme case except the $g_{vv}$ component which becomes $\mathcal{N}(\zeta,\bar{\zeta})\rho^{2}+\mathcal{O}(\rho^{3})$. The displacement memory with this consideration can be achieved by setting $\kappa=0$ in Eq.(\ref{ngd3n}), and written as
\begin{align}
\Delta S^{\bar{\zeta}}_{E} =\frac{\rho}{4}(1+\zeta\bar{\zeta})^{2}(\Delta\lambda_{\zeta\bar{\zeta}} S^{\bar{\zeta}}_{E}+\Delta\lambda_{\zeta\zeta} S^{\zeta}_{E})+\mathcal{O}(\rho^{2}), \label{ngd2}
\end{align}
where subscript $E$ stands for the change in the displacement vector for \textit{extremal} or \textit{zero temperature} configuration. One can again take the variation of $\lambda_{AB}$ along Killing direction and obtain set of differential equations similar to the fixed-temperature analyses. The related details can be found in Bhattacharjee et al\cite{Bhattacharjee2021}.% It is difficult to have explicit solutions of supertranslation or superrotation which can give rise the same shift or change in $\lambda_{AB}$ sitting in $\Delta S^{\bar{\zeta}}_{E}$.

%\paragraph{Relation with BMS symmetry:}
%Here also, we commented on the possibility of obtaining the BMS symmetry that induces the shift in GDE taking $f=T+vX$. In principle, it can be determined from (\ref{chi2}), by choosing $\lambda_{AB}$ linear in $v$. It is difficult to obtain an explicit solution for $X$ or $Y$. Therefore, we provided the full analysis for reduced form of the metric.

\subsubsection{Memory \& BMS symmetries: for a less generic form of the metric}
%\textbf{\large Memory for Reduced Metric:} \\

In this section, we consider a less generic form of the full metric (\ref{m}) by setting $g_{vA}=0$ which can be regarded as an asymptotic form of a metric near the horizon of a spherically symmetric black hole deformed in the spatial sector. The displacement memory for this setup can be computed in similar way as obtained for the full metric. 

We find that the fixed temperature configuration does not produce very interesting result. However, the displacement memory is written in terms of $\Delta\lambda_{AB}$, and further set of conditions can be obtained in order to have the connection with BMS symmetries. The interesting finding appears if we consider zero temperature configuration. The change in the displacement vector is written in terms of $\Delta\lambda_{AB}$. Using the $vv$-component of Einstein field equation, the displacement memory is given by
\begin{align}
    \Delta S^{\bar{\zeta}}_{E} =\frac{\rho}{4}(1+\zeta\bar{\zeta})^{2}H(\zeta,\bar{\zeta}) \Delta v S^{\bar{\zeta}}_{E}+\mathcal{O}(\rho^{2}). \label{mxt4d}
\end{align}
We notice that the displacement memory is proportional to $\rho$ whereas in the far region case\cite{strominger2018lectures} the memory is proportional to $\frac{1}{r}$. This implies that the displacement memory near the horizon of black holes is mimicking the one obtained null infinity. 

One can further relate it with the BMS symmetries. It turns out that if we freeze off the superrotation, we have an exact solution: $X(\zeta,\bar{\zeta})=X_{1}(\zeta)+X_{2}(\bar{\zeta})$. Hence, there is a  supertranslation $X(\zeta,\bar{\zeta})$ that can induce the same shift in the displacement memory. We have also considered the three dimensional analyses for extreme and non-extreme cases, and its connection with near-horizon BMS symmetries in Bhattacharjee et al\cite{Bhattacharjee2021}. %Next, we conclude our findings below by providing a brief summary.

%We notice crucial  differences  between  memory  effect  and  BMS  symmetries  obtained  for the far region and near the horizon of a black hole.  The  extra  set  of supertranslation  generator  does not get  recovered  at  the  null  infinity.   Further difference seems to be in the structure of GDE. The near-horizon GDE contains a linear derivative term of the deviation vector together with double derivative with respect to time, whereas the far region GDE contains only a double-time derivative of the deviation vector to the leading order. The function $\lambda$( or $\lambda_{AB}$ in four dimensions) mimics the derivative of News tensor, $C_{\zeta\zeta}$, in the far region. These differences make the  analyses  of  near-horizon  memory effect significantly  non-trivial  than  the  far  region  case. %This is our complete analysis of displacement memory effect and its connection with BMS symmetries near the horizon of three and four dimensional extremal and non-extremal black holes.

\section{Discussion and Outlook}\label{discuss}
The primary motivation of this article is to provide a review study on the displacement memory effect near the horizon of black holes and its connection with asymptotic symmetries which is similar to the one established at null infinity for asymptotically flat spacetimes. In this respect, we started with a brief introduction of the intrinsic formulation of null shell. As a result, we show that the supertranslation-like transformations can be achieved in the context of soldering of two black hole spacetimes. We explicitly provided the results for Schwarzschild and ERN cases. However, interestingly, superrotation can not be recovered for ERN spacetime whereas it can be recovered for Schwarzschild. Further, as a result of interaction between horizon shell and test particles, we find that the particles remain on the initial 2-dimensional surface for Schwarzschild consideration as surface current is zero whereas for ERN spacetime, particles get displaced off from the initial 2-dimensional surface. We compute the components of deviation vectors which carry the supertranslation parameter ensuring that the memory can be obtained in terms of BMS parameters. 

We have provided a detailed description of GW memory effect near the horizon of black holes as a more realistic approach which is analogous to the one obtained at null infinity. As a major distinguishing feature in order to establish a connection with asymptotic symmetries, we observed that there are two supertranslation parameters $T(x^{A})$ and $X(x^{A})$ and one superrotation $Y^{A}(x^{A})$ in near horizon analyses whereas there is only one supertranslation in the far region case. We also notice that $\Delta\lambda_{AB}$ is the data available to be considered in the detection which mimics the data $C_{zz}$ present near the null infinity\cite{strominger2018lectures}. The form of the GDE is also quite different with respect to the far region case. These are the brief and major distinguishing features between displacement memory obtained near the horizon of black holes and at asymptotic null infinity, together with its possible connection to BMS symmetries.

%Researchers actively involved in this field are expecting to observe the GW-memory experimentally in the next decade. 
The observational features of GW will be extremely useful in investigating the signatures of the asymptotic symmetries in displacement memory effect. As a result, we might be able to look into such symmetries in greater detail as a firm evidence in near future. The theoretical aspects of our study might help as a model framework in the detection prospects of the near-horizon BMS memory effect. It is expected that present-day detectors like LIGO might not be able to play a crucial role in the detection prospects. We hope that advanced detectors like aLIGO or LISA might be able to capture this effect as LISA is looking for a much longer wavelength opening up the detection realm to a wider range of gravitational wave sources. In this direction, building up the theoretical framework of post-Newtonian (PN) formalism having relevance for asymptotic symmetries, it is interesting to investigate the contribution of supertranslation-like symmetries in non-oscillatory signals of the gravitational wave polarizations. This would surely provide a more direct approach to gravitational wave data analysts in order to search for asymptotic symmetries in GW memory signal.%The implementation of wave extraction techniques to carry out the simulations for measuring the non-linear GW memory generated by binary black hole mergers \cite{Pollney_2011} is equally important to explore.

As an alternative approach, at first, it seems that detecting astrophysical signatures of asymptotic symmetries from  gravitational  lensing  might  be  an  appropriate approach. However, any supertranslated  geometry  of  the spacetime will not lead any deviations to the standard results in general relativity \cite{Comp_re_2016}. The underlying reason to the problem is that given  a supertranslated Schwarzschild black hole, one can always choose a coordinate patch in a finite region where the metric will be given  by  the  Schwarzschild metric. However, we still have a hope to detect such symmetries in black hole shadows and lensing if we have a dynamically evolving spacetime carrying supertranslation field. So this might set a stronger grounds for detecting the asymptotic symmetries through deflection angle approach and black hole shadows. Interestingly, it is not difficult to understand that why gravitational memory is suitable for detecting such symmetries, because it appears as a physical effect where initial and final vacua differ by a BMS supertranslation. On another hand, the post-Newtonian tidal environment analysis especially in terms of BMS symmetries can also be explored from BMS-detection prospects. %tidal heating where late in the inspiral, individuals of abinary are highly sensitive to each other’s tidal fields as the bodies approach their final plunge and merger. 

Furthermore, as we have been investigating the issues from classical perspectives, the quantum memory effect has also got considerable attention in very recent, and it is yet to be explored extensively. This would certainly give a new meaning to the quantum treatment of the memory to catch on to the hawking information paradox. We know that the displacement memory is induced by the radiative energy flux, and it has been shown that there exists a new kind of gravitational memory- \textit{spin  memory  effect} which is sourced by angular momentum flux\cite{Pasterski2016}. It is interesting to examine the signatures of asymptotic symmetries in the context of spin memory effect near the horizon of black holes from BMS-detection point of view. Further, the algebra of asymptotic symmetries on null surface situated at a finite location of the manifold might determine some fascinating role of symmetries on GW memory. 

\section{Acknowledgement}
I would like to express my sincere thanks to Srijit Bhattachrajee for proofreading this article and sharing his stimulating ideas during the preparation of the draft. I would also like to thank Subhodeep Sarkar and Arpan Bhattacharyya for illuminating discussions. The support from DST-SERB, Government of India under the scheme Early Career Research Award (File no.: ECR/2017/002124) through the project titled \textit{Near
Horizon Structure of Black Holes} is also acknowledged. 

\bibliography{references}

\begin{thebibliography}{10}

\bibitem{PhysRevLett.116.061102}
B.~P. Abbott, R.~Abbott and e.~a. Abbott, Observation of gravitational waves
  from a binary black hole merger, {\em Phys. Rev. Lett.} {\bf 116}, p. 061102
  (Feb 2016).

\bibitem{PhysRevLett.125.101102}
R.~Abbott, T.~D. Abbott and e.~a. Abraham, Gw190521: A binary black hole merger
  with a total mass of $150\text{ }\text{ }{M}_{\ensuremath{\bigodot}}$, {\em
  Phys. Rev. Lett.} {\bf 125}, p. 101102 (Sep 2020).

\bibitem{Zeldovich:1974gvh}
Y.~B. Zel'dovich and A.~G. Polnarev, {Radiation of gravitational waves by a
  cluster of superdense stars}, {\em Sov. Astron.} {\bf 18}, p.~17  (1974).

\bibitem{Braginsky:1985vlg}
V.~B. Braginsky and L.~P. Grishchuk, {Kinematic Resonance and Memory Effect in
  Free Mass Gravitational Antennas}, {\em Sov. Phys. JETP} {\bf 62}, 427
  (1985).

\bibitem{Christodoulou:1991cr}
D.~Christodoulou, {Nonlinear nature of gravitation and gravitational wave
  experiments}, {\em Phys. Rev. Lett.} {\bf 67}, 1486  (1991).

\bibitem{PhysRevD.89.084039}
L.~Bieri and D.~Garfinkle, Perturbative and gauge invariant treatment of
  gravitational wave memory, {\em Phys. Rev. D} {\bf 89}, p. 084039 (Apr 2014).

\bibitem{Strominger:2014pwa}
A.~Strominger and A.~Zhiboedov, {Gravitational Memory, BMS Supertranslations
  and Soft Theorems}, {\em JHEP} {\bf 01}, p. 086  (2016).

\bibitem{PhysRevD.101.124010}
A.~A. Rahman and R.~M. Wald, Black hole memory, {\em Phys. Rev. D} {\bf 101},
  p. 124010 (Jun 2020).

\bibitem{Favata:2011qi}
M.~Favata, {The Gravitational-wave memory from eccentric binaries}, {\em Phys.
  Rev. D} {\bf 84}, p. 124013  (2011).

\bibitem{PhysRevD.80.024002}
M.~Favata, Post-newtonian corrections to the gravitational-wave memory for
  quasicircular, inspiralling compact binaries, {\em Phys. Rev. D} {\bf 80}, p.
  024002 (Jul 2009).

\bibitem{Favata_2009}
M.~Favata, {Nonlinear} {gravitational}-{wave} {memory} {from} {binary} {black}
  {hole} {mergers}, {\em The Astrophysical Journal} {\bf 696}, L159 (apr 2009).

\bibitem{Favata_2009n}
M.~Favata, Gravitational-wave memory revisited: Memory from the merger and
  recoil of binary black holes, {\em Journal of Physics: Conference Series}
  {\bf 154}, p. 012043 (mar 2009).

\bibitem{Islo:2019qht}
K.~Islo, J.~Simon, S.~Burke-Spolaor and X.~Siemens, {Prospects for Memory
  Detection with Low-Frequency Gravitational Wave Detectors} (6 2019).

\bibitem{Lasky:2016knh}
P.~D. Lasky, E.~Thrane, Y.~Levin, J.~Blackman and Y.~Chen, {Detecting
  gravitational-wave memory with LIGO: implications of GW150914}, {\em Phys.
  Rev. Lett.} {\bf 117}, p. 061102  (2016).

\bibitem{Boersma:2020gxx}
O.~M. Boersma, D.~A. Nichols and P.~Schmidt, {Forecasts for detecting the
  gravitational-wave memory effect with Advanced LIGO and Virgo}, {\em Phys.
  Rev. D} {\bf 101}, p. 083026  (2020).

\bibitem{Pollney_2011}
D.~Pollney and C.~Reisswig, {Gravitational} {memory} {in} {binary} {black}
  {hole} {mergers}, {\em The Astrophysical Journal} {\bf 732}, p. L13 (apr
  2011).

\bibitem{Grant:2021hga}
A.~M. Grant and D.~A. Nichols, {Persistent gravitational wave observables:
  Curve deviation in asymptotically flat spacetimes} (9 2021).

\bibitem{PhysRevD.101.023011}
M.~H\"ubner, C.~Talbot, P.~D. Lasky and E.~Thrane, Measuring gravitational-wave
  memory in the first ligo/virgo gravitational-wave transient catalog, {\em
  Phys. Rev. D} {\bf 101}, p. 023011 (Jan 2020).

\bibitem{Islam:2021old}
T.~Islam, S.~E. Field, G.~Khanna and N.~Warburton, {Survey of gravitational
  wave memory in intermediate mass ratio binaries} (9 2021).

\bibitem{Hubner:2021amk}
M.~H\"ubner, P.~Lasky and E.~Thrane, {Memory remains undetected: Updates from
  the second LIGO/Virgo gravitational-wave transient catalog}, {\em Phys. Rev.
  D} {\bf 104}, p. 023004  (2021).

\bibitem{doi:10.1098/rspa.1962.0161}
H.~Bondi, M.~G.~J. Van~der Burg and A.~W.~K. Metzner, Gravitational waves in
  general relativity, vii. waves from axi-symmetric isolated system, {\em
  Proceedings of the Royal Society of London. Series A. Mathematical and
  Physical Sciences} {\bf 269}, 21  (1962).

\bibitem{strominger2018lectures}
A.~Strominger, Lectures on the infrared structure of gravity and gauge theory
  (2018).

\bibitem{PhysRevLett.116.231301}
S.~W. Hawking, M.~J. Perry and A.~Strominger, Soft hair on black holes, {\em
  Phys. Rev. Lett.} {\bf 116}, p. 231301 (Jun 2016).

\bibitem{PhysRevD.96.064013}
P.-M. Zhang, C.~Duval, G.~W. Gibbons and P.~A. Horvathy, Soft gravitons and the
  memory effect for plane gravitational waves, {\em Phys. Rev. D} {\bf 96}, p.
  064013 (Sep 2017).

\bibitem{PhysRevLett.116.091101}
L.~Donnay, G.~Giribet, H.~A. Gonz\'alez and M.~Pino, Supertranslations and
  superrotations at the black hole horizon, {\em Phys. Rev. Lett.} {\bf 116},
  p. 091101 (Mar 2016).

\bibitem{Donnay2016}
L.~Donnay, G.~Giribet, H.~A. Gonz{\'a}lez and M.~Pino, Extended symmetries at
  the black hole horizon, {\em Journal of High Energy Physics} {\bf 2016}, p.
  100 (Sep 2016).

\bibitem{Blau:2015nee}
M.~Blau and M.~O'Loughlin, {Horizon Shells and BMS-like Soldering
  Transformations}, {\em JHEP} {\bf 03}, p. 029  (2016).

\bibitem{PhysRevD.98.104009}
S.~Bhattacharjee and A.~Bhattacharyya, Soldering freedom and
  bondi-metzner-sachs-like transformations, {\em Phys. Rev. D} {\bf 98}, p.
  104009 (Nov 2018).

\bibitem{PhysRevD.43.1129}
C.~Barrab\`es and W.~Israel, Thin shells in general relativity and cosmology:
  The lightlike limit, {\em Phys. Rev. D} {\bf 43}, 1129 (Feb 1991).

\bibitem{nullsurface}
C.~Barrabes and P.~A. Hogan, {\em Singular Null Hypersurfaces in General
  Relativity: Light-Like Signals from Violent Astrophysical} (World Scientific
  Publishing Co. Pte. Ltd., 2003).

\bibitem{PhysRevD.100.084010}
S.~Bhattacharjee, S.~Kumar and A.~Bhattacharyya, Memory effect and bms-like
  symmetries for impulsive gravitational waves, {\em Phys. Rev. D} {\bf 100},
  p. 084010 (Oct 2019).

\bibitem{PhysRevD.102.044041}
S.~Bhattacharjee and S.~Kumar, Memory effect and bms symmetries for extreme
  black holes, {\em Phys. Rev. D} {\bf 102}, p. 044041 (Aug 2020).

\bibitem{PhysRevLett.105.111103}
G.~Barnich and C.~Troessaert, Symmetries of asymptotically flat
  four-dimensional spacetimes at null infinity revisited, {\em Phys. Rev.
  Lett.} {\bf 105}, p. 111103 (Sep 2010).

\bibitem{barnich2012supertranslations}
G.~Barnich and C.~Troessaert, Supertranslations call for superrotations
  (2012).

\bibitem{Volonteri_2005}
M.~Volonteri, P.~Madau, E.~Quataert and M.~J. Rees, The distribution and cosmic
  evolution of massive black hole spins, {\em The Astrophysical Journal} {\bf
  620}, 69 (feb 2005).

\bibitem{Gou_2014}
L.~Gou, J.~E. McClintock, R.~A. Remillard, J.~F. Steiner, M.~J. Reid, J.~A.
  Orosz, R.~Narayan, M.~Hanke and J.~Garc{\'{\i}}a, {Confirmation} {via} {the}
  {continuum}-{fitting} {method} {that} {the} {spin} {of} {the} {black} {hole}
  {in} {Cygnus} x-1 {is} {extreme}, {\em The Astrophysical Journal} {\bf 790},
  p.~29 (jun 2014).

\bibitem{McClintock_2006}
J.~E. McClintock, R.~Shafee, R.~Narayan, R.~A. Remillard, S.~W. Davis and L.-X.
  Li, The spin of the near-extreme kerr black hole {GRS} 1915+105, {\em The
  Astrophysical Journal} {\bf 652}, 518 (nov 2006).

\bibitem{Brenneman_2006}
L.~W. Brenneman and C.~S. Reynolds, Constraining black hole spin via x-ray
  spectroscopy, {\em The Astrophysical Journal} {\bf 652}, 1028 (dec 2006).

\bibitem{Strominger1996}
A.~Strominger and C.~Vafa, Microscopic origin of the bekenstein-hawking
  entropy, {\em Physics Letters B} {\bf 379}, 99 (Jun 1996).

\bibitem{CARTER1966423}
B.~Carter, The complete analytic extension of the reissner-nordström metric in
  the special case e2 = m2, {\em Physics Letters} {\bf 21}, 423  (1966).

\bibitem{PhysRevD.62.024005}
S.~Liberati, T.~Rothman and S.~Sonego, Nonthermal nature of incipient extremal
  black holes, {\em Phys. Rev. D} {\bf 62}, p. 024005 (Jun 2000).

\bibitem{Bhattacharjee2021}
S.~Bhattacharjee, S.~Kumar and A.~Bhattacharyya, Displacement memory effect
  near the horizon of black holes, {\em Journal of High Energy Physics} {\bf
  2021}, p. 134 (Mar 2021).

\bibitem{PhysRevD.98.124016}
L.~Donnay, G.~Giribet, H.~A. Gonz\'alez and A.~Puhm, Black hole memory effect,
  {\em Phys. Rev. D} {\bf 98}, p. 124016 (Dec 2018).

\bibitem{Comp_re_2016}
G.~Comp{\`{e}}re and J.~Long, Classical static final state of collapse with
  supertranslation memory, {\em Classical and Quantum Gravity} {\bf 33}, p.
  195001 (sep 2016).

\bibitem{Pasterski2016}
S.~Pasterski, A.~Strominger and A.~Zhiboedov, New gravitational memories, {\em
  Journal of High Energy Physics} {\bf 2016}, p.~53 (Dec 2016).

\end{thebibliography}
\bibliographystyle{ws-procs961x669}
%\bibliography{ws-pro-sample}

\end{document}